\documentclass[
    ,final            
  ]
  {aipproc}

\layoutstyle{6x9}

\newcommand{\bmath}{\begin{displaymath}}
\newcommand{\emath}{\end{displaymath}}
\newcommand{\be}{\begin{equation}}
\newcommand{\ee}{\end{equation}}
\newcommand{\bea}{\begin{eqnarray}}
\newcommand{\eea}{\end{eqnarray}}

\newcommand{\gf}{$G_F$}

\begin{document}

\title{ Radiative Corrections and the Universality of 
the Weak Interactions   }

\author{Alberto Sirlin}{
  address={Department of Physics, New York University, \\ 
4 Washington Place, New York, NY 10003, USA}
}

%

\begin{abstract}
We review the radiative corrections to superallowed Fermi transitions and neutron
$\beta$ decay in the Standard Model, and their relevance for the universality
of the Weak Interactions.
%

\end{abstract}

\maketitle


The study of radiative corrections (R.C.) to $\mu$ and $\beta$ decays has 
played an important role in the analysis of weak interactions before and after 
the emergence of the Standard Model (SM).

In the framework of the local V-A theory that preceded the SM, the R.C. to  
$\mu$ decay are convergent, to first order in \gf and all orders in $\alpha$,
after charge and mass renormalization \cite{c1}. 
The corrections to the electron
spectrum in $\mu$ decay are large and play an important role in verifying that
the parameters $\rho$ and $\delta$ equal $3/4$, a major prediction 
of the V-A theory \cite{c2}.

The expression for the $\mu$ lifetime is
\be
\frac{1}{\tau_\mu} = \frac{G_F^2\ m_\mu^5}{192 \pi^3}\ f
\left(\frac{m_e^2}{m_\mu^2}\right)\left[ 1 + \delta_\mu  \right]\, , 
\ee
where $f(x) = 1 - 8x - 12 x^2\ \mbox{ln}x + 8 x^3 - x^4$ and $\delta_\mu$ is
the R.C.. One finds $\delta_\mu = - 4.1995 \times 10^{-3} + 1.5 \times 10 ^{-6} + \ldots$,
where the first and second terms are the ${\cal O}(\alpha)$ and 
${\cal O}(\alpha^2)$ contributions \cite{c2, c3}.
This leads to
$G_F = 1.16637(1) \times 10^{-5}/ \mbox{GeV}^2$.

Instead, the R.C. to $\beta$ decay in the V-A theory are 
logarithmically divergent. For some time it was thought that form factor
effects from the strong interactions (S.I.) can give rise to an effective 
cutoff. However, using current algebra (C.A.) techniques, Bjorken, and Abers,
Dicus, Norton, and Quinn \cite{c4} studied the short distance behavior of 
the R.C. to $\beta$ decay and reached the conclusion that the S.I. cannot tame
their logarithmic divergence!

In the SM with three generations, the interactions of $W^{\pm}$ with 
fermions are given by
\bmath
-(g_0/\sqrt{2}) W_\mu \left[ \overline{U} \gamma^\mu a_{-} D' +
\overline{N}' \gamma^\mu a_{-} L \right] + \mbox{h.c.}\, ,
\emath
where $a_{-}=(1-\gamma_5)/2$, $D' = VD$, $V$ is the unitary $3\times 3$
CKM matrix, and U, D, N, and L are column vectors representing the 
up and down quarks, the neutrinos, and the charged leptons. The
 principle of non-abelian gauge invariance tells us that $g_0$ 
is a universal parameter, independent of the nature of the fermions involved. 
That fundamental property of  $g_0$ and the unitarity of $V$ may be regarded as
the present statement of universality.

Since the SM is renormalizable, it should provide a convergent answer for the 
R.C. to $\beta$ decay! In fact, using a simplified version of the SM,
and neglecting the S.I., it was found in 1974 that, to very good approximation,
the corrections are the same as in the V-A theory, 
with $\Lambda \to M_Z$! \cite{c5} During 1974-1978 the Current Algebra 
Formulation was extended to the real SM, including the effect of the 
S.I.\cite{c6}. This leads to the following expression for pure Fermi 
 $\beta$ decay transitions:
\be \label{eq5}
{\cal P} d^3 p = {\cal P}^0 d^3 p \left\{ 1 + \frac{\alpha}{2 \pi} \left[
3 \mbox{ln}\left(\frac{M_Z}{m_p}\right) + g(E,E_m) + 6 \overline{Q}\ 
\mbox{ln}\left(\frac{M_Z}{M}\right) + 2C + {\cal A}_{\overline{g}} \right]
\right\}\, ,
\ee
\be
{\cal P}^0 d^3 p = \frac{G_F^2\ (V_{ud})^2}{8 \pi^4} |M_F|^2\ F(Z,E)\
(E_m-E)^2\ d^3 p\, .
\ee
The first two terms between square brackets in Eq.~(\ref{eq5}) arise from
the vector current and are independent of the S.I..
In fact, the proton mass $m_p$ cancels in the sum. 
The function  $g(E,E_m)$, where $E$ is the energy of the electron 
or positron and $E_m$ its end-point energy,
describes the R.C. to the spectrum in $\beta$ decay 
in the presence of S.I.. 
It was first derived using the so-called ``$1/k$'' method \cite{c7}.

The third term between the square brackets in Eq.~(\ref{eq5}) is a short distance 
contribution to the Fermi amplitude arising from the axial vector current.
$\overline{Q}$ is the average charge of the fundamental doublet involved in the transition.
In the SM this is the $u$-$d$ doublet and we have $\overline{Q}=(2/3-1/3)/2=1/6$.
The $2C$ term is a corresponding non-asymptotic part while ${\cal A}_{\overline{g}} \sim -0.34$ 
is a very small asymptotic contribution from QCD.

The R.C. to $\beta$ decay are dominated by a large logarithmic term:
$(3\alpha/2\pi)\mbox{ln}(M_Z/2E_m)$. As an example, for the superallowed
$^{14}O$ Fermi transition $E_m = 2.3 \mbox{MeV}$, and this correction amounts to
$3.4\%$. It turns out that such a large correction is phenomenologically
crucial to verify the unitarity of the CKM matrix. Early smoking gun for the
SM at the level of the quantum corrections?

Contributions
of ${\cal O}(Z \alpha^2)$ and ${\cal O}(Z^2 \alpha^3)$ are denoted by 
$\delta_2$ and $\delta_3$. One finds that $\delta_2$
varies from $0.22\%$ for $^{14}O$ to $0.50\%$ for $^{54}Co$, while $\delta_3$ 
is much smaller \cite{c8}.

There is also a correction $\delta_c$ that reflects the lack of perfect overlap
between the wavefunctions of the parent and daughter nuclei due to
Coulomb forces and configuration mixing effects in the shell-model wavefunctions.
It has been extensively discussed in the literature \cite{c9,c10}.

Leading logarithms of ${\cal O}\left[(\alpha/\pi)\mbox{ln}(M_Z/m_p)\right]^n$
$(n \ge 2)$ have been incorporated by means of a renormalization group analysis \cite{c11}.

Putting these various contributions together, and integrating over the positron momentum
one obtains
\bmath
\Gamma = \Gamma^0 \left\{1+\frac{\alpha(m_p)}{2\pi}\left[\overline{g}(E_m)+{\cal A}_{\overline{g}}
\right]+\frac{\alpha}{2\pi}\left[\mbox{ln}\left(\frac{m_p}{M}\right)+ 2C \right]
+\delta_2+\delta_3 \right\}
\emath
\be \label{eq9}
\times\ S(m_p,M_Z) (1-\delta_c) \, ,
\ee
where $S(m_p,M_Z) = 1.0225$ is the short distance contribution and $\overline{g}(E_m)$
is the average of  $g(E,E_m)$ over the positron spectrum. The term
$(\alpha/2\pi)\left[\mbox{ln}(m_p/M) +2C\right]$ from the axial vector current
is model dependent.
In recent discussions the mass $M$, that represents the onset of the 
asymptotic  behavior, is allowed to vary in the range $m_{A1}/2 \le M \le 2m_{A1}$,
with a central value $M_c = m_{A1} = 1.26 \mbox{GeV}$, the mass of the
$A_1$ resonance, which has the correct quantum numbers to mediate 
that contribution \cite{c9,c12}. 
Jaus and Rasche proposed to split $C = C_{Born}+C_{NS}$,
where the first term is identified with the Born approximation calculation of the diagram where 
the insertions of the axial vector and electromagnetic currents involves the same
nucleon, while $C_{NS}$ corresponds to the contributions in which
the insertions occur in different ones \cite{c13}. One obtains 
$C_{Born} = 0.881 \pm 0.030$.

In order to verify CVC, it is advantageous to factor out the nuclear-dependent part of
the R.C.. A simple way of doing this is to factor out the expression in Eq.~(\ref{eq9})
in the form $(1+\delta_R)(1+\Delta_R)(1-\delta_c)$, where
\be
1+\delta_R = 1+\frac{\alpha(m_p)}{2\pi}\overline{g}(E_m)+\frac{\alpha}{\pi}C_{NS}+
\delta_2+\delta_3\, ,
\ee
\be
1+\Delta_R =\left\{ 1+\frac{\alpha}{2\pi} \left[  \mbox{ln}\left(\frac{m_p}{M}\right)
+ 2 C_{Born} +{\cal A}_{\overline{g}}  \right] \right\} S(m_p,M_Z) \, .
\ee
One can then introduce a radiatively corrected ${\cal F} t$ value
\be
{\cal F} t = f t (1+\delta_R)(1-\delta_c)\ = K / 2 {G'}_V^2\, ,
\ee
\be
K = 2 \pi^3 \mbox{ln}2\, \, \hbar^7 / m_e^5 c^4 = 8.12027 \times 10^{-7} (\hbar c)^6
\mbox{GeV}^{-4}s\, ;
\ee
\be
{G'}_V^2 = {G}_V^2  (1+\Delta_R)\, \, ;\, \, G_V = G_F V_{ud} \, .
\ee
The test of CVC consists in checking the constancy of the ${\cal F} t$ values.
Using then the average ${\cal F} t$, one obtains ${G'}_V^2$. Inserting
the calculated $\Delta_R$ one obtains $G_V^2$ and therefore $V_{ud}$.
A recent determination by Towner and Hardy is 
$V_{ud}=0.9740 \pm 0.0005$ \cite{c9} (nuclear $\beta$ decay).
It is important to note that the error is mainly theoretical ($\pm 4 \times 10^{-4}$ 
from $\Delta_R$, $\pm 3 \times 10^{-4}$ from $\delta_c$).

In the case of neutron $\beta$ decay, we avoid nuclear physics complexities, 
but this is not a pure Fermi transition!
However, we can apply C.A. in combination with the $1/k$ method \cite{c7}. The latter allows
the calculation of some important observables in the presence of the S.I. in terms
of effective coupling constants ${G'}_V$ and  ${G'}_A$, neglecting small contributions
of ${\cal O}\left((\alpha/\pi)(E/M)\mbox{ln}(M/E), (\alpha/\pi)(q/M)\right)$, where $M$ is a
hadronic mass. The observables include the correction to the electron spectrum (given by
$(\alpha/2\pi)g(E,E_m)$), the longitudinal polarization of electrons, and the electron asymmetry
from polarized neutrons.

We use the  $1/k$ method to express the lifetime and the electron asymmetry
in terms of  ${G'}_V$ and  ${G'}_A$. The inverse lifetime is proportional 
to  ${G'}_V^2 +3 {G'}_A^2$. The asymmetry gives us ${G'}_A/{G'}_V$. Combining
the two observables, we can find  ${G'}_V$. Using  ${G'}_V^2 = G_F^2 V_{ud}^2(1+\Delta_R)$,
we extract  $V_{ud}$.

Employing ${G'}_A/{G'}_V = -1.2690 \pm 0.0022$ and $\tau_n = 885.6 \pm 0.8\ \mbox{s}$,
a recent analysis by Towner and Hardy \cite{c9} gives $|V_{ud}| = 0.9745 \pm 0.0016$ 
(neutron $\beta$ decay), which is consistent with the nuclear result but has considerably
larger error. Combining with  $|V_{us}| = 0.2196 \pm 0.0026$ and  
$|V_{ub}| = 0.0036 \pm 0.0007$, recommended by PDG02 \cite{c14}, one obtains
\be \label{eq10}
\sum_i |V_{ui}|^2 = 0.9969 \pm 0.0015 \quad \mbox{(nuclear $\beta$ decay \cite{c10})}\, ,
\ee
\be
\sum_i |V_{ui}|^2 = 0.9979 \pm 0.0033 \quad \mbox{(neutron $\beta$ decay \cite{c11})}\, .
\ee
The first test is short by $2.1 \sigma$, while the second one is in agreement, but has a larger error.
On the other hand, PDG02 averages only over recent asymmetry experiments with polarization
 $> 90 \%$, leading to ${G'}_A/{G'}_V = -1.2720 \pm 0.0018$ and $|V_{ud}| = 0.9725 \pm 0.0013$
(neutron) \cite{c14}, and a $2.2 \sigma$ shortfall.

Based on a recent high statistics experiment \cite{c15}, a preliminary value 
$\mbox{Br}\ (K^+ \to \pi^0 e^+ \nu) = (5.13\pm0.2\pm0.08\pm0.04)\%$ has been reported,
which is higher than the PDG02 entry $(4.82\pm0.06)\%$. If the result is confirmed and 
the lifetime is not modified, it may lead to a solution of the unitarity deviation. In fact,
the central value in the unitarity test would become
$(0.9740)^2+(0.2196)^2\, \, 5.13/4.82 = 1.000002$! 
Of course, it would be important to check the experimental status of $K^0 \to \pi^- e^+ \nu_e$,
as well as the $K_{\mu 3}$ modes.

It is also interesting to remember that the deviation in Eq.~(\ref{eq10}) can be removed in
``manifest'' left-right symmetric models \cite{c16} by choosing
$2 \zeta = 0.0031 \pm 0.0015$, where $\zeta$ is the mixing angle \cite{c9,c12}.

The determination of $V_{us}$ is derived mainly from $K_{l3}$ decays applying R.C.
and chiral perturbation theory (ChPT). One considers $K^+ \to \pi^0 e^+ \nu$, 
$K^0 \to \pi^- e^+ \nu$, and  $K_{\mu 3}$ modes.
After applying R.C. the experiments determine $f_+(0) V_{us}$. To get  $V_{us}$,
we need  $f_+(0)$.
For $K^0 \to \pi^- l^+ \nu$, the non-renormalization theorem tells us that  $f_+(0)$
differs from 1 by terms of second order in the mass splittings \cite{c16b}.
Expanding $f_+(0) = 1 + f_1 + f_2 + \cdots$, where $f_1 = {\cal O}(m_q\ \mbox{ln}m_q)$,
 $f_2 = {\cal O}(m_q^2\ \mbox{ln}m_q)$, and $m_q$ are generic quark masses,  
$f_1$ was obtained in a model independent manner \cite{c17} and lowers  $f^{K^0 \pi^-}_+(0)$ to
$0.977$, while an estimate for  $f_2$ gives $f_2 = -0.016 \pm 0.008$ \cite{c18}.
Combining the two results, one has  $f^{K^0 \pi^-}_+(0) = 0.961 \pm 0.008$ \cite{c18}. 
For  $K^+ \to \pi^0 l^+ \nu$, there is a complication. One finds 
$|\pi^0> = \cos\epsilon |3> + \sin\epsilon |8>$ where  $|\pi^0>$ is the physical
state and $\epsilon = (\sqrt{3}/4)(m_d - m_u)/(m_s - \hat{m}) \approx 0.01$.
As a consequence, to zeroth order in $m_q$, there is a breaking of isospin invariance
and $f_{K^+ \pi^0}(0)/f_{K^0 \pi^-}(0) = 1.0172$.
Including terms of  ${\cal O}(\epsilon m_q)$, the ratio becomes 1.022 \cite{c17,c18}.
Thus, there is an interesting isospin breaking effect that enhances the $K^+ \to \pi^0 e^+ \nu$
rate by $4.45 \%$ relative to $K^0 \to \pi^- e^+ \nu$.
Using the above results, the experimental data, some of which had been corrected by 
long distance R.C., and including the short distance R.C., Leutwyler and Roos obtained
$|V_{us}| = 0.2196 \pm 0.0023$ \cite{c18}, while PDG02 recommends  
$|V_{us}| = 0.2196 \pm 0.0026$ \cite{c14}.

Very recently, the R.C. to $K_{l3}$ decays have been studied in the ChPT framework \cite{c19},
leading to $|V_{us}| = 0.2201 \pm 0.0024$, very close to the other determinations.

Also very recently Bijnens and Talavera have discussed the evaluation of $K_{l3}$ decays to
two-loop order in  ChPT in the isospin limit \cite{c20}. Their expression for $f_+(0)$
depends on two unknown constants that can in principle be determined by accurate
measurements of the scalar form factor $f_0(t) = f_+(t) + f_-(t) t/(M_K^2 - M_\pi^2)$,
specifically its slope and curvature.

Bill Marciano tells me that precise lattice calculations of $f_+(0)$ are possible.
Lattice practitioners should be encouraged to carry out this important calculation!
  
\bigskip

This work was supported in part by NSF Grant PHY-0245068.







\end{document}